\documentclass[11pt]{article} 

\usepackage[utf8]{inputenc} 

\usepackage{authblk}

\usepackage{geometry} 
\usepackage{graphicx} 

\usepackage{amssymb}
\usepackage{amsmath}
\usepackage{amsthm}
\usepackage{mathtools}
\usepackage{stmaryrd}
\usepackage{xfrac}

\usepackage{booktabs} 
\usepackage{array} 
\usepackage{paralist} 
\usepackage{verbatim} 
\usepackage{subfig} 
\usepackage{makecell}

\usepackage{fancyhdr} 
\usepackage{sectsty}
\allsectionsfont{\sffamily\mdseries\upshape} 

\usepackage[nottoc,notlof,notlot]{tocbibind} 
\usepackage[titles,subfigure]{tocloft} 

\title{T- Hop: A framework for studying the importance  path information in molecular graphs for chemical property prediction.  }

\author[1]{Abdulrahman Ibraheem}
\author[1,2]{Narsis Kiani}
\author[1]{Jesper Tegner}

\affil[1]{KAUST, Saudi Arabia.}
\affil[2]{Karolinska Institute, Sweden.}
\date{}                     

\begin{document}

\maketitle

\begin{abstract}
This paper studies the usefulness of incorporating path information in predicting chemical properties from molecular graphs, in the domain of QSAR (Quantitative Structure-Activity Relationship). Towards this, we developed a GNN-style model which can be toggled to operate in one of two modes: a non-degenerate mode which incorporates path information, and a degenerate mode which leaves out path information. Thus, by comparing the performance of the non-degenerate mode versus the degenerate mode on relevant QSAR datasets, we were able to directly assess the significance of path information on those datasets. Our results corroborate previous works, by suggesting that the usefulness of path information is dataset-dependent. Unlike previous studies however, we took the very first steps towards building a model that could predict  upfront whether or not path information would be useful for a given dataset at hand. Moreover, we also found that, albeit its simplicity, the degenerate mode of our model yielded rather surprising results, which outperformed more sophisticated SOTA models in certain cases. 
\end{abstract}

\section{Introduction}
Here, we study the importance of graph path information in predicting chemical properties from molecular graphs, a topic which falls under the topic of QSAR in the field of cheminformatics. For that purpose, we developed a framework called T-Hop which allows us to incorporate path information into a graph neural network (GNN) -style model. To allow us study the effect of path information, the framework comprises two modes: a non-degenerate mode which incorporates path information; and a degenerate mode which does not use path information. We applied the framework to six datasets from the MoleculeNet suite of cheminformatics datasets \cite{molecule_net}. Results suggest that the effect of path information on accuracy is dataset-dependent. Also, we found no strong (Pearson) correlation between model performance and the maximum path length used in the model. Yet, we found that a degenerate case of our model yields positively surprising results on four out of the six datasets studied.

The present work's application domain falls under the field of cheminformatics. Currently, the field of cheminformatics is witnessing great interest within the research community. This is due to a number factors. First, cheminformatics has immense utility in the real world. For example denovo drug design \cite{drug_design_2024} \cite{drug_design_maksym}, retrosynthesis frameworks \cite{retrosynthesis_chemirise}, QSAR (Quantitative Structure-Activity Relationship) models \cite{qsar}, as well as molecular similarity and sub-structure matching tools \cite{structural_similarity_search},  can relieve chemists and their parent pharmaceutical companies of a significant portion of the pain, energy, time and costs involved in the drug discovery and design process. Second, progress in the field of cheminformatics has been further fast-tracked by three favorable factors: 1) the increased availability of relevant cheminformatics data \cite{chem_data_pubchem_1} \cite{chem_data_tox21} \cite{chem_data_sider}  \cite{chem_data_chembl}  \cite{chem_data_schneider}; 2) the great strides recorded in machine learning, especially deep learning, which affords an avenue to extract useful information from the available data; and 3). the increased availability of compute and storage resources which supports the physical realization of the preceding two factors. 

More specifically, this work focuses on datasets related to the branch of cheminformatics called QSAR \cite{qsar}.  QSAR is guided by the intuition that the observed physical, toxicological, physiological, and bioactivity-related properties of molecules are informed  by the molecules' underlying structures \cite{carhart}. Approaches under QSAR include:  1). prediction of a target variable, such as aqeous solubility, from a compendium of calculable molecular descriptors \cite{molecular_descriptor_padel} \cite{molecular_descriptor_6} \cite{molecular_descriptor_7} \cite{molecular_descriptor_8}; 2). the use of 2-d fingerprints \cite{fingerprint_1} \cite{fingerprint_2} \cite{fingerprint_3} as inputs to a neural network or any other machine learning classifier; 3). the use of graph neural networks, which typically use primitive atomic features as inputs, with optional additional use of edge features and 3-d geometry information.  \cite{coley_attrib} \cite{molecule_net} \cite{attentivefp} \cite{weave} 4); the use of smiles string as inputs in analogy with NLP models \cite{safe} \cite{chem_bert_a} \cite{mol_former}; and 5). more recently, the use of the 2-d drawings/schematics of molecules \cite{img_mol}, drawing inspiration from the field of computer vision. 

Our T-Hop framework falls within the genre of graph neural networks (GNNs). For QSAR, GNNs represent molecules as graphs whose nodes are the atoms of the molecules, and whose edges are the bonds of the molecules. A unifying theme of most GNN-based models is that they compute neighborhood-aware representations for each node-centered neighborhood of the graph. In the context of molecules, this means computing embeddings for molecular sub-structures comprised of each atom and its neighboring atoms and bonds. 
The task of computing neighborhood-aware embeddings in GNNs has been approached in various ways in the literature. Generally, models differ in how they aggregate and combine information in the neighborhood of each node/edge. For example, some models use only node information \cite{earliest_gnn} \cite{gcn}, whereas others blend node information with edge information \cite{mpnn} \cite{dmpnn}\cite{coley_attrib}. Also, some models use an attention mechanism \cite{gat} \cite{gat_v2}, while others jettison it  altogether. Our T-Hop  uses two pieces of information: node features and a linear combination of 2-d slices of a 3-d matrix containing path information.

\section{Related work}
T-Hop, being a GNN-style framework which incorporates path information, bears resemblance to several exiting GNNs that either implicitly or explicitly use path information. First, our framework is similar to IGCN \cite{igcn} in the sense that the latter uses a powered symetrically normalised identity-shifted adjacency matrix (PSNIA), which can ultimately be expressed in terms of the powered adjacency matrix, which in turn is known to contain path information. Morever, the results of \cite{igcn} are quite similar to ours, because in their experiments, utilizing path information does not yield better results in all cases. Further, Mix-Hop \cite{mix_hop} also resembles our model because it also uses path information via the PSNIA. In fact, Mix-Hop  involves a more extensive use of path information than IGCN, because it uses a set of PSNIAs with different powers, corresponding to different path lengths. Each layer of the model concatenates embeddings associated with different path lengths. While experiments in the Mix-Hop paper suggested that path information boosts accuracy, a more extensive set of experiments might ultimately show otherwise. Further, our framework also bears similarity to Power-Up \cite{power_up} because the latter uses a matrix that records information about the path length of the shortest distance between all pairs of nodes in the graph in question. Similarly to \cite{mix_hop}, they consider a set of matrices, each associated with a specific  shortest distance path length. However, unlike \cite{mix_hop}, they do not  concatenate information across the matrices. Rather, similar to our approach, they use a learnable linear combination of the matrices. 

Further, a line of works   \cite{geodesic_gnn} \cite{path_net} \cite{path_nn} \cite{path_rnn}   explicitly considers the sequence of paths leading to the nodes of a graph. In some cases, all paths are considered \cite{path_rnn}, while in some cases, shortest paths are considered \cite{path_nn}, while in yet other cases, paths are sampled to lighten computational costs \cite{path_net}. For example, in PathNet \cite{path_net}, path information is harnessed, with the aim of capturing graph global stucture, especially in heterophily networks. They first sample paths using a maximal entropy path sampler, which takes the graph's structure into consideration. Then using an LSTM-like cell, they compute node embeddings for target nodes by inputing the sequence of nodes on the target node's path into the cell. To make the model distance-aware, a separate set of weights is learnt for each relevant distance. Experimental results showed that PathNet works better for heterophily graphs compared to homophily ones. Further, Geodesic graph neural nets (GDGNN) \cite{geodesic_gnn} seeks to imporove on the expressiveness of GNNs, by incorporating so-called geodesic information, which essentially distills to pooling information along  paths in the graph.  In the GDGNN approach, a conventional GNN is first applied to the graph to compute embeddings for each node.  So-called horizontal and vertical geodesics are then computed along relevant paths of the graph, by pooling embeddings along those paths. In the case of vertical geodesics, node degree and distance information are concatenated with the embeddings before pooling. A possible drawback is that the path associated with the horizontal pooling is chosen at random, to dodge high computational costs. However, this can lead to a situation wherein the pooling layer outputs different results for the same graph at diferent times. Moreover, experiments showed that GDGNN does not always outperform baseline methods. More recently, the authors of PathNN\cite{path_nn} experimented with all paths between all pairs of nodes as well as with shortest paths, and put forth a theory that states that the set of all paths is more discriminatory than the 1-WL test. Similar to PathNet, they used an LSTM for modelling paths as distance-aware  sequences of nodes (and edges). Their experiments showed that incorporating path information does not lead to better results in all cases.  

Two aspects of Graphormer \cite{graphormer} also fit the above paradigm. First, in Graphormer, the shortest distance between every pair of nodes is used to augment the attention matrix. Second, the edges along the shortest path joing every pair of nodes is considered. Then the average of the dot products between each edge and a position-aware set of weights is computed and further used to augment the attention matrix. The above paradigm has also been extended to the domain of knowledge base completion. For example, in \cite{path_rnn}, following a compositional approach, they consider all paths between a source entity and a destination entity. For each path, they use an RNN to model the sequence of relations encountered from the source node to the destination node, with the goal of computing an output embedding that captures the direct relation between the source entity and the destination entity. A drawback of Path-RNN is that it considers all possible paths and this leads to huge computational costs .

Other works incorporating path information include NBFNet (Neural Bellman-Ford Networks) \cite{bellman_ford_net} and its special cases  \cite{katz_index} \cite{distance_edit} \cite{page_rank} . Inspired by the Bellman-Ford algorithm \cite{bellman_ford}, NBFNet  considers all paths between every pair of nodes in a graph. To alleviate computational complexity, they treat the paths in parallel, treating each path as a product of edges along the path. To cast the formulation unto a GNN setting, they used neural networks to: model the product of edges as a message passing function; and model the sum of  paths as an aggregation function. Distinct advantages of NBFNet includes its suitability for both the transductive and inductive settings, as well as its modest computational complexity.  However, unlike our framwework, NBFNet focuses on link prediction rather than graph classification.

\section{ Introducing the T-Hop Framework}
\label{intro_t_hop}

 This section describes T-Hop, the framework we developed for studying the importance of path information for the domain of chemical property prediction from molecular graphs. Let $G= (V, E)$ represent a graph, where $V = \{v_1,...,v_n\}$ and $E = \{e_1, ..., e_m\}$ are the nodes and edges of $G$, as usual. The adjacency matrix of $G$ is $A$. We consider two arbitrary nodes, $v_i$ and $v_j $ in graph $G$. Let $\mathcal{B}^{L}_{i,j,k}$  be the number of paths of length $L$ between $v_i$ and $v_j$ that contain $v_k$. Clearly, we can arrange the values, $\mathcal{B}^{L}_{i,j,k}$, in an $n \times n \times n$ 3-d tensor denoted $\mathcal{B}^{L} $ $\in \mathbb{R}^{n\times n \times n}$, such that the entry on the $i$-th row, $j$-th column and $k$-th depth of $\mathcal{B}^{L}$ is  $\mathcal{B}^{L}_{i,j,k}$. Using  $\mathcal{B}^{L}_{i,j,k}$, we now define a 3-d tensor, $\mathcal{T}^{L}$  $\in \mathbb{R}^{n\times n \times n}$, as a scaled version of  $\mathcal{B}^{L}$ .   Specifically, where $\mathcal{T}^{L}_{i,j,k}$ is the entry on the $i$-th row, $j$-th column and $k$-th depth of $\mathcal{T}^{L}$, we define $\mathcal{T}^{L}_{i,j,k}$ as follows:

\begin{equation}
\label{eqn_1}
\mathcal{T}^{L}_{i,j,k} = \frac{ \mathcal{B}^{L}_{i,j,k} } {(L+1)}
\end{equation}
 
To proceed, we define a new $n \times n$ matrix $\mathcal{M}$ as follows:
\begin{equation}
\label{eqn_2}
\mathcal{M} =   \alpha_0 A + \sum_{L = 2}^{L_{m}}\sum_{k = 0}^{n-1} \alpha_{L,k}\mathcal{T}^{L}_{:,:,k}
\end{equation}

Above,  $A$ is adjacency matrix as usual, while $\alpha_0$ and $\alpha_{L,k}$ are learnable scalars. Further, $L_m$ is the maximum path length considered in the model, chosen due to computational considerations. Also, $\mathcal{T}^{L}_{:,:,k}$  $\in \mathbb{R}^{n\times n} $ is the  2-d matrix slice associated with depth-$k$ of $\mathcal{T}^{L}$.  We note from the  above formula that $\mathcal{M}$ is simply a learnt linear combination of the adjacency matrix, $A$, and depth-$k$ slices of tensor $\mathcal{T}^{L}$. Using $\mathcal{M}$,  the $l$-th layer of our T-Hop model can now be described as:

\begin{equation}
\label{eqn_3}
H^{l+1} =  \sigma( \mathcal{M} \: H^{l}W^{l} )
\end{equation}

Above, $ H^{l} $ denotes input features to the $l$-th layer, and $W^l$ denotes a learnable matrix of weights, and $\sigma(.)$ is a non-linear activation function. As usual, for the first layer of the network, we define $H^{l}= X$ where $X \in \mathbb{R}^{n\times d}$ is the set of $d$-dimensional  input features associated with the $n$ nodes of the graph. Going back to Equation \ref{eqn_2}, we see that when the second term on the R.H.S of the equation is zero, we simply have  $\mathcal{M} =   \alpha_0 A $. When we use this value of  $\mathcal{M}$ in Equation \ref{eqn_3}, viz a viz $H^{l+1} =  \sigma( \mathcal{M} \: H^{l}W^{l} )$, we call the resulting model a  \lq \lq degenerate model."  Importantly, it should be emphasized that the degenerate model does not use any path information at all. Rather, it only uses the adjacency matrix. Indeed, a  centerpiece of this work is to empirically contrast the degenerate model with the non-degenerate model.    

\section{ Connection between  $\mathcal{T}^{L}$ and the powered adjacency matrix.}

On the one hand, the $(i, j)$ entry of the powered adjacency matrix, $A^{L}$ contains the number of paths of length $L$ between nodes $v_i$ and $v_j$, while on the other hand,  the $(i, j, k)$ entry of  $\mathcal{B}^{L}$  contains the number of paths of length $L$ between $v_i$ and $v_j$ passing through $v_k$. Based on this, it is intuitive that a relationship should exist between $A^{L}$ and $\mathcal{B}^{L}$.  Moreover, given that $\mathcal{T}^{L} = \frac{ \mathcal{B}^{L}} {(L+1)}$, it is also intuitive that a relationship should exsit between ${A}^{L}$  and $\mathcal{T}^{L}$. We now explore the connection between  ${A}^{L}$  and $\mathcal{T}^{L}$ as follows. We start by defining the function, $f_{sum}:\mathbb{R}^n \rightarrow \mathbb{R}$, which simply outputs the sum of all components of  its input vector. It can be shown that we would obtain the powered adjacency matrix, $A^L$, when we apply $f_{sum}$ to $\mathcal{T}^{L}$. We first give the following definition:

\newtheorem{def1}{Definition}[section] 
\begin{def1}[Cardinality of multiset $\mathcal{P}^L$ ]
\label{def_1}
Let $G$ be a graph of nodes, $V = \{v_1,..., v_n\}$, and edges. Let  $v_i$ and $v_j$ be any two arbitrary nodes in $G$, and Let $A^L_{ij}$ be the number of simple paths of length $L$ between $v_i$ and $v_j$. Let $ P_q^L = \{ v_1^q, v_2^q,..., v_{L+1}^q\}$ denote the $q$-th simple path of  length $L$  between $v_i$ and $v_j$, where $ v_k^q$ is the $k$-th node in the $q$-th path, $P_q^L$. Let  $\mathcal{P}^L = \{P_1^L, P_2^L, ... P_{ A_{ij}}^L \} = \{ v_1^1, v_2^1, ..., v_{L+1}^1,  \;\; v_1^2, v_2^2, ..., v_{L+1}^2, \;\; ..., \;\; v_1 ^{A_{ij}}, v_2 ^{A_{ij}}, ..., v_{L+1} ^{A_{ij}} \} $ be a \textbf{multiset} containing all the simple paths of  length $L$  between $v_i$ and $v_j$. \textbf{Then, the cardinality of multiset $\mathcal{P}^L$  is defined as the number of nodes in $\mathcal{P}^L$, counting multiplicities of nodes. }
\end{def1}

\noindent Based on the preceding definition, the following is a fact:

\newtheorem{thm1}[def1]{Fact }
\begin{thm1}[Cardinality of multiset $\mathcal{P}^L$  equals $\sum_k \mathcal{B}^{L}_{i,j,k}$ ]
\label{thm_1}
 \textbf{The cardinality of multiset $\mathcal{P}^L$ defined in Definition \ref{def_1} above is equal to $\sum_k \mathcal{B}^{L}_{i,j,k}$}
\end{thm1}

\begin{figure}
\includegraphics[height = 120pt, width =200pt]{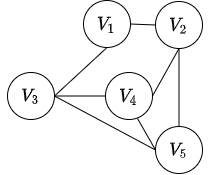} 
\caption{An illustrational graph of five nodes }
\label{fig_1}
\end{figure}

\begin{proof}
 The proof is best sketched with an example. As an example, let us use the graph $G$ of five nodes depicted in Figure \ref{fig_1}. Without loss of generality, let us consider all simple paths of length $L = 3$ between two arbitrary nodes, $v_1$ and $v_5$ in the graph. From the graph, we see that there are 2 simple paths of length $3$ between $v_1$ and $v_5$, so that $A_{1,5}^3 = 2$. These paths are $P_1^3 = \{v_1, v_2, v_4, v_5\}$ and  $P_2^3 = \{v_1, v_3, v_4, v_5\}$. Hence, we may write 
$\mathcal{P}^3 = \{P_1^3, P_2^3\} = \{v_1, v_2, v_4, v_5 \;\; v_1, v_3, v_4, v_5\}$. Upon sorting $\mathcal{P}^3$, we now have $\mathcal{P}^3 = \{v_1, v_1,  \;\; v_2,\;\; v_3, \;\; v_4, v_4,  \;\; v_5, v_5 \}$. Now, given any node, $v_k$ in $G$, when we count the multiplicity of $v_k$ in the sorted version of $\mathcal{P}^3$, we see it corresponds to the number of simple paths of length $3$ between $v_1$ and $v_5$ that contain $v_k$. For example, we see clearly that node $v_4$ has multiplicity of 2, because it is contained in two different paths of length 3 between $v_1$ and $v_4$, whereas node $v_2$ has multiplicity of $1$, because it is contained in a single path of length 3 between $v_1$ and $v_5$. Generalizing this observation, we see that, if $\mathcal{P}^L$ is the multiset of nodes that constitute the simple paths of length $L$ between any two arbitrary nodes, $v_i$ and $v_j$, then for any  node, $v_k \in \mathcal{P}^L$ , the multiplicity of $v_k$ in $\mathcal{P}^L$ corresponds to the number of paths of length $L$ between $v_i$ and $v_j$ that contain $v_k$, which in turn, by definition, is equal to  $\mathcal{B}^{L}_{i,j,k}$. \textbf{In summary, for any  node, $v_k \in \mathcal{P}^L$, the multiplicity of $v_k$ in $\mathcal{P}^L$ equals $\mathcal{B}^{L}_{i,j,k}$}.  Based on this, we now consider the quantity $\sum_k \mathcal{B}^{L}_{i,j,k}$. It should be clear that the quantity $\sum_k \mathcal{B}^{L}_{i,j,k}$ simply equals the sum of multiplicities of all nodes in  $\mathcal{P}^L$, which, in turn, equals the cardinality of $\mathcal{P}^L$.
\end{proof}

\noindent Next, using  Fact \ref{thm_1}, we have the following proposition: 

\newtheorem{thm2}[def1]{Proposition } 

\begin{thm2}[$f_{sum}$ recovers $A^L$ from  $\mathcal{T}^{L}$  ]
Let $f_{sum}:\mathbb{R}^n \rightarrow \mathbb{R}$ be the function that takes a vector $u \in \mathbb{R}^n$ as input and returns as output the summation of all components of $u$. Then, with  $ t^L_{ij}$ denoting the $n$-dimensional vector that stretches along the depth-axis of the $3$-d tensor, $\mathcal{T}^{L}$, at a given $i$-th row, $j$-th column position of $\mathcal{T}^{L}$, we have that $f_{sum}(t_{ij}^L) = A^L_{ij}$. 
\end{thm2}
\begin{proof}

To proceed, from Definition \ref{def_1} above, we recall  the meaning of the multiset $\mathcal{P}^L = \{P_1^L, P_2^L, ... P_{ A_{ij}}^L\}   =  \{ v_1^1, v_2^1, ..., v_{L+1}^1,  \;\; v_1^2, v_2^2, ..., v_{L+1}^2, \;\; ..., \;\; v_1 ^{A_{ij}}, v_2 ^{A_{ij}}, ..., v_{L+1} ^{A_{ij}} \}$; we also bring to mind the definition of the multiset's cardinality given therein. In particular, the number of paths in $\mathcal{P}^L$ is $A_{ij}$ and each path contains  $L + 1$ nodes, so that the cardinality of $\mathcal{P}^L$ is equal to  $(L+1)A_{ij}^L$.  Hence, we have $|\mathcal{P}^L| = (L+1)A_{ij}^L$. But, we already know from  Fact \ref{thm_1} above that  $ |\mathcal{P}^L|  = \sum_k \mathcal{B}^{L}_{i,j,k}$.  Hence, we have $ \sum_k \mathcal{B}^{L}_{i,j,k} = (L+1)A_{ij}^L$, implying $ \sum_k \dfrac{ \mathcal{B}^{L}_{i,j,k} }{(L+1)} = A_{ij}^L $. Further, by definition, we know $\dfrac{ \mathcal{B}^{L}_{i,j,k} }{(L+1)} = \mathcal{T}^{L}_{i,j,k}$. Thus, $\sum_k \mathcal{T}^{L}_{i,j,k} = A_{ij}^L$. Now, it is clear that $ \sum_k \mathcal{T}^{L}_{i,j,k}$ is tantamount to applying $f_{sum}$ to  the vector $  t^L_{ij} = \mathcal{T}^{L}_{i,j,k}$, which completes the proof.
\end{proof}

Proposition 4.3 reveals that T-Hop should be at least as expressive as a model that uses the sum of the adjacency matrix, $A$ and the powered adjacency matrices , $A^L$. To see this, we begin with what the proposition says: $ A_{ij}^L  = \sum_k \mathcal{T}^{L}_{i,j,k} \Longrightarrow$  $\sum_{L=2}^{L_m} A_{ij}^L = \sum_{L=2}^{L_m} \sum_k \mathcal{T}^{L}_{i,j,k} \implies$   $\sum_{L=2}^{L_m} A^L = \sum_{L=2}^{L_m} \sum_k \mathcal{T}^{L}_{:,:,k} $.  Now,  if we restrict T-Hop by setting all the learnable parameters in Equation \ref{eqn_2} to unity (i.e. setting  $\alpha_0 = 1$, and  $\alpha_{L,k}= 1$ for all $L$ and $k$) then  Equation \ref{eqn_2} distills to $\mathcal{M} =   A +  \sum_{L = 2}^{L_{m}}\sum_k \mathcal{T}^{L}_{:,:,k}$, which makes it clear that $\mathcal{M} = A + \sum_{L=2}^{L_m} A^L $ under the restrictive setting that  $\alpha_0 = 1$, and  $\alpha_{L,k}= 1$ for all $L$ and $k$. Hence, by removing this restriction, T-Hop is expected to be more expressive.

\section{Experiments}
We performed experiments in the domain of molecular property prediction \cite{molecule_net} \cite{compt} \cite{grover}. By far, the most popular benchmark datasets in this domain are the MoleculeNet suite of datasets popularized by \cite{molecule_net}. These datasets span both regression and classification tasks on the one hand, as well as single-task and multi-task classification problems on the other hand. Also, the authors of \cite{molecule_net} considered four ways of splitting the datasets: random splitting, scaffold splitting, stratified splitting, and time splitting. In addition,  some authors have also introduced a form of scaffold splitting called balanced scaffold splitting (e.g. \cite{compt}). For each dataset, the authors of \cite{molecule_net}, recommended a particular way of splitting the dataset in question. For example, for the ClinTox dataset \cite{molecule_net}, they recommend random splitting, whereas for the BBBP dataset \cite{molecule_net}, they recommend scaffold splitting. However, a more recent trend in the literature favors the jettisoning of the random splitting method, for the adoption of the scaffold splitting method. This trend is due to how the scaffold splitting method is better able to assess the generalization strengths of the proposed model/method. This is because the scaffold splitting method strives to make the molecules in the test set to be as different as possible from those in the training set. Thus,  we focussed on the scaffold splitting method. Also, throughout our experiments, we adopted the standard 80:10:10 splitting for the training/validation/test partitions of the datasets.
We performed experiments on six datasets from the MoleculeNet suite of datasets. Three of the datasets (ClinTox, BACE and BBBP ) involve classification tasks, while  the other three (FreeSolv, ESOL and Lipophilicity) invlove regression tasks.  For the classification tasks, the pertinent metric is AUC-ROC, so that the higher the better, whereas for the regression tasks, the pertinent metric is R.M.S.E so that the lower the better. For each dataset, using a combination of automated hyper-parameter search programs (such as TPE (Tree Parzen Estimator) from Optuna \cite{optuna} and ASHA (Asynchronous Successive Halving Algorithm) from Ray Tune \cite{ray_tune} ), as well as hand-tuning, we searched for optimal hyper-parameters on the validation portion of the dataset. We chose the epoch that yielded best result on the test set. Following previous work, we ran randomly initialized models three times on each dataset, and recorded the mean and standard deviation of the runs. Our experiments were aimed at exploring four research questions. First, to see whether or not incorporating path information into our framework can yield significant gains in accuracy on the aforementioned datasets. Second, to see if there is a relationship between the maximum path length, $L_m$ and accuracy. Third, given a dataset, to see if it is possible to predict upfront whether or not path information can yield gains in accuracy on the dataset. Fourth, to see how the framework compares with SOTA methods.

\subsection{ Juxtaposition of degnerate case against non-degenerate case and relationship between accuracy and maximum path length}

It can be recalled from Section \ref{intro_t_hop} that the T-Hop framework has two modes: 1). a non-degenerate mode which incorporates path information, and which corresponds to $L_m > 1$, where $L_m$ is maximum path length; and 2) a degenerate mode which does NOT incorporate path information, and which corresponds to $L_m = 1$. Consequently, to explore whether or not incorporating path information into the framework leads to better performance or not, we juxtaposed the degenerate case with the non-degenerate case . We present the results in Table \ref{tab:degen_vs_non_degen}, where we see that path information increases accuracy on two out of the six datasets being considered. This suggests that incorporating path information does not always boost accuracy, and that, generally, the capability of path information to boost accuracies is dataset-dependent. Indeed, this observation aligns with experiments from some previous works \cite{igcn}. For example, in the IGCN  paper \cite{igcn}, experiments on the AWA2 dataset \cite{awa2_dataset} showed worse performance as $k$ increased from $1$ to $3$, where $k$ can be viewed as a parameter that controls the maximum path length incorporated into the IGCN model. Likewise, in PathNN \cite{path_nn} whereas increasing path length improved results on the NCI1 dataset of the TUDataset collection \cite{tu_dataset}, it failed to improve results on the PROTEINS and ENZYMES datasets of the same TUDataset collection. Moreover, even in recent work on dynamic graphs,  viz GraphMixer \cite{graph_mixer} it was found that using a larger receptive field, and hence path length, diminished accuracies. 

Further, we also conducted an experiment where we observed accuracies as $L_m$ increased from $1$ through $5$ on each of the six datasets. We show results in Table \ref{tab:all_powers}. For each dataset, the last column of the table shows the Pearson correlations between performance measures and $L_m$. For the classification datasets (first three datasets), a high positive correlation would mean that incorporating path information boosts performance, while for the regression datasets (last three datasets) a high negative correlation would mean the same. However, as can be seen, we don't observe any such correlation values in the last column of the table, showing that increasing path length does not correlate strongly with performance boosts.

 \begin{table}
\caption{Juxtaposition of Results for degenerate case ($L_m = 1$) against non-degenerate case ($L_m > 1$)}
\label{tab:degen_vs_non_degen}
\centering
\begin{tabular}{cccc}
\toprule

Dataset&Metric &\makecell{Result for \\$L_m = 1$}  & \makecell{Best Result for\\$L_m > 1$}  \\ 
\hline \hline
BACE&     ROC-AUC  $\uparrow$  &          $\mathbf{86.4_{(0.003)}}$  &  $ 82.1_ {(0.003)}$     \\ 
BBBP&     ROC-AUC  $\uparrow$ &          $\mathbf{73.5_{(0.003)}}$  &  $ 70.0_ {(0.037)}$     \\ 
ClinTox&     ROC-AUC $\uparrow$   &          $91.2_{(0.017)}$  &  $ \mathbf{91.8_ {(0.015)}}$     \\ 

\hline 
&&&\\
\hline \hline
FreeSolv&     R.M.S.E  $\downarrow$ &          $\mathbf{1.93_{(0.132)}}$  &  $ 1.97_ {(0.035)}$     \\ 
ESOL&     R.M.S.E   $\downarrow$  &          $\mathbf{0.90_{(0.019)}}$  &  $ 0.96_ {(0.017)}$     \\ 
Lipophilicity&     R.M.S.E  $\downarrow$   &          $0.74_{(0.016)}$  &  $\mathbf{ 0.71_ {(0.010)}}$     \\ 
\end{tabular}
\newline\newline

\end{table}

 \begin{table}
\footnotesize
\caption{Model performance accross all values of $L_m$}
\label{tab:all_powers}
\centering
\begin{tabular}{cccccccc}
\toprule

Dataset  &  Metric &  $L_m = 1$  &   $L_m = 2$   &  $L_m = 3$ &   $L_m = 4$&   $L_m = 5$  &\makecell{Corr-\\elation} \\ 
\hline \hline
BACE&     $\uparrow$  &          ${86.4}$  &  $ 81.9$  & $ 82.1$&  $ 80.1$  &$ 77.6$     &$-0.95$     \\ 
BBBP&     $\uparrow$  &          ${73.5}$  &  $ 67.9$  & $ 69.9$&  $ 70.0$  &$ 67.2$      &$-0.68$    \\ 
ClinTox&     $\uparrow$  &          ${91.2}$  &  $ 88.1$  & $ 89.6$&  $ 89.6$  &$ 91.8$    &$0.29$     \\ 

\hline 
&&&&&&\\
\hline \hline

FreeSolv&     $\downarrow$  &          ${1.93}$  &  $ 1.97$  & $ 2.46$&  $ 2.71$  &$ 2.86$      &$0.97$    \\ 
ESOL&     $\downarrow$  &          ${0.90}$  &  $ 0.93$  & $ 0.96$&  $ 1.00$  &$ 1.03$            &$0.99$ \\ 
Lipophilicity&     $\downarrow$  &          ${0.74}$  &  $ 0.71$  & $ 0.71$&  $ 0.71$  &$ 0.74$   &$0.12$    \\ 
\end{tabular}
\end{table}

\subsection{Towards upfront prediction of when path information helps}
We now turn to the question of: given a dataset, can we contrive a way/rule to predict upfront whether or not path information would help on it ? To motivate this question, we consider the compute times associated with computing $A$ for $L_m = 1$, as well as $\mathcal{T}$ and $A$ for $L_m > 1$. We display the pertinent training compute times  in Table \ref{tab:training_times}. As expected the compute times increase as $L_m$ increases. Considering this increase in compute times with $L_m$, it becomes important to ask: given a specific dataset, should we expend the extra computational effort of using $L_m > 1$ on it or not ? Clearly, if we could contrive a rule to predict upfront that incorporating path information on a given dataset would not help, then we could save ourselves the extra compute time associated with using path information on that dataset, by simply using the degenerate mode of T-Hop. To forge ahead, we fell back on classical graph properties (e.g. graph diameter, closeness centrality, density, average clustering etc.). The idea is to use these properties as features in a machine learning classifier to make the aforementioned desired prediction. For each of the six datasets in this work, we computed fifteen such graph properties. However, because for some of the graph properties, we needed to compute: 1) means of node properties acrosss the dataset  2) standard deviations of node property means across the dataset   3) mean of node property standard deviations across dataset, we ultimately arrived at a total of 36 features for each dataset. We show the computed properties in Tables \ref{tab:props_1_to_5},  \ref{tab:props_6_to_10} and  \ref{tab:props_11_to_15}. Based on the preceding, for each of the six datasets, we computed a 36-dimensional feature vector. Hence, we got just 6 samples; each sample is 36-d vector associated with one of the six datasets. Towards building our predictor/classifier, we used the 3 samples associated with the 3 regression datasets as training samples, and used the remaining 3 samples as test samples.  More specifically, we built a binary classifier trained to output 1 if the dataset associated with the input vector benefits from path information; otherwise, the classifier should output 0. Testing on the 3 test samples showed the classifier can successfully make accurate predictions on two of the three cases, thereby achieving a prediction accuracy of $66.67\%$. Thus, we see, given more datasets to train on, it is possible in principle to build a classifier that can predict upfront whether path information would be helpful on a given dataset. Of equal importance, this classification result also suggests that the classical graph properties used as features into the classifier offer significant explanation for why certain datasets benefit from path information, whereas others do not.

 \begin{table}
\footnotesize
\caption{Single epoch training compute times in secs. across all values of $L_m$}
\label{tab:training_times}
\centering
\begin{tabular}{ccccccc}
\toprule

Dataset  &      $L_m = 1$  &   $L_m = 2$   &  $L_m = 3$ &   $L_m = 4$&   $L_m = 5$  \\ 
\hline \hline

BACE &             $1.09$  &  $ 10.13$  & $ 22.09$&  $ 35.54$  &$ 50.69$         \\ 
BBBP&              $1.35$  &  $ 27.58$  & $ 67.23$&  $ 103.90$  &$ 150.96$        \\ 
ClinTox&          $1.13$  &  $ 25.39$  & $ 51.55$&  $ 82.06$  &$ 120.12$      \\ 

\hline 
&&&&&&\\
\hline \hline

FreeSolv&            ${0.59}$  &  $ 0.70$  & $ 0.82$ &$ 0.93$  &$ 1.09$        \\ 
 ESOL&                 $0.76$  &  $ 1.74$  & $ 3.21$&  $ 5.35$  &$ 7.59$           \\ 
Lipophilicity&        ${2.25}$  &  $ 44.97$  & $ 91.48$&  $ 148.76$  &$ 214.57$     \\ 

\end{tabular}
\end{table}

 \begin{table}

\caption{First five graph properties across all six datasets }
\label{tab:props_1_to_5}
\centering
\begin{tabular}{ccccccc}
\toprule
Dataset  &  \makecell[t]{Path\\Info\\Helped ?} &  \makecell[t]{Node\\Degree}          &  \makecell[t]{Max\\Weight\\Clique} &\makecell[t]{ Diam-\\eter}     & \makecell[t]{ Density}                                &  \makecell[t]{Modu-\\larity}  \\
\midrule
FreeSolv&         NO                                         & $\makecell[t]{1.83\\0.28\\0.67}$  &   $\makecell[t]{2.00\\0.11}$           &  $\makecell[t]{5.04\\2.11}$  &   $\makecell[t]{0.30\\0.16}$    &  $\makecell[t]{0.30\\0.14}$  \\
\midrule
ESOL&         NO                                         & $\makecell[t]{1.98\\0.24\\0.72}$  &   $\makecell[t]{2.01\\0.12}$           &  $\makecell[t]{7.02\\3.35}$  &   $\makecell[t]{0.22\\0.13}$    &  $\makecell[t]{0.39\\0.14}$  \\
\midrule
Lipophilicity&         YES                                         & $\makecell[t]{2.18\\0.06\\0.69}$  &   $\makecell[t]{2.05\\0.22}$           &  $\makecell[t]{13.85\\4.04}$  &   $\makecell[t]{0.09\\0.03}$    &  $\makecell[t]{0.60\\0.08}$  \\

\midrule
BACE&         NO                                         & $\makecell[t]{2.17\\0.06\\0.76}$  &   $\makecell[t]{2.10\\0.30}$           &  $\makecell[t]{15.22\\3.37}$  &   $\makecell[t]{0.07\\0.02}$    &  $\makecell[t]{0.65\\0.05}$  \\

\midrule
BBBP&         NO                                         & $\makecell[t]{2.13\\0.14\\0.76}$  &   $\makecell[t]{2.03\\0.16}$           &  $\makecell[t]{11.28\\4.18}$  &   $\makecell[t]{0.11\\0.07}$    &  $\makecell[t]{0.54\\0.10}$  \\

\midrule
ClinTox&         YES                                       & $\makecell[t]{2.10\\0.17\\0.77}$  &   $\makecell[t]{2.03\\0.18}$           &  $\makecell[t]{12.38\\6.01}$  &   $\makecell[t]{0.12\\0.09}$    &  $\makecell[t]{0.55\\0.13}$  \\
\end{tabular}
\end{table}

 \begin{table}
\caption{Next five graph properties across all six datasets }
\label{tab:props_6_to_10}
\centering
\begin{tabular}{ccccccc}
\toprule
Dataset  &  \makecell[t]{Path\\Info\\Helped ?} &  \makecell[t]{Shortest\\Path\\Length}          &  \makecell[t]{Closeness\\Centrality}       &\makecell[t]{ Between-\\ness\\centrality}     & \makecell[t]{Edge\\ Betw.\\centrality}      & \makecell[t]{ Eigen-\\vector\\centrality}   \\
\midrule
FreeSolv&         NO                                         & $\makecell[t]{2.46\\0.74}$                              &   $\makecell[t]{0.46\\0.15\\0.08}$           &  $\makecell[t]{0.23\\0.07\\0.22}$              &   $\makecell[t]{0.35\\0.14\\0.08}$            &  $\makecell[t]{0.35\\0.11\\0.11}$  \\
\midrule
ESOL&         NO                                         & $\makecell[t]{3.15\\1.13}$                                   &   $\makecell[t]{0.37\\0.14\\0.07}$           &  $\makecell[t]{0.21\\0.06\\0.20}$              &   $\makecell[t]{0.28\\0.12\\0.09}$            &  $\makecell[t]{0.28\\0.10\\0.11}$  \\
\midrule
LIPOPHILICITY&         YES                                         & $\makecell[t]{5.61\\1.40}$                   &   $\makecell[t]{0.20\\0.06\\0.04}$           &  $\makecell[t]{0.18\\0.02\\0.17}$              &   $\makecell[t]{0.19\\0.02\\0.13}$            &  $\makecell[t]{0.17\\0.03\\0.09}$  \\
\midrule

BACE&         NO                                         & $\makecell[t]{6.29\\1.30}$                              &   $\makecell[t]{0.17\\0.04\\0.03}$           &  $\makecell[t]{0.17\\0.02\\0.18}$              &   $\makecell[t]{0.17\\0.02\\0.14}$            &  $\makecell[t]{0.14\\0.02\\0.10}$  \\
\midrule
BBBP&         NO                                         & $\makecell[t]{4.70\\1.46}$                              &   $\makecell[t]{0.24\\0.09\\0.05}$           &  $\makecell[t]{0.18\\0.04\\0.17}$              &   $\makecell[t]{0.20\\0.06\\0.12}$            &  $\makecell[t]{0.19\\0.06\\0.10}$  \\
\midrule
ClinTox&        YES                                         & $\makecell[t]{5.13\\2.15}$                              &   $\makecell[t]{0.24\\0.11\\0.05}$           &  $\makecell[t]{0.18\\0.04\\0.18}$              &   $\makecell[t]{0.21\\0.08\\0.12}$            &  $\makecell[t]{0.19\\0.08\\0.10}$  \\

\end{tabular}
\end{table}

 \begin{table}

\caption{Last five graph properties across all six datasets }
\label{tab:props_11_to_15}
\centering
\begin{tabular}{ccccccc}
\toprule
Dataset  &  \makecell[t]{Path\\Info\\Helped ?} &  \makecell[t]{clustering coefficient}          &  \makecell[t]{Smallest\\Laplacian\\eigenvalue} &\makecell[t]{ Second\\Smallest\\Laplacian\\eigenvalue}     & \makecell[t]{ Second\\Larglest\\Laplacian\\eigenvalue}    
&  \makecell[t]{Largest\\Laplacian\\eigenvalue}  \\
\midrule
FreeSolv&         NO                                         & $\makecell[t]{0.003\\0.04\\0.002}$  &   $\makecell[t]{-6.26\\4.21e^{-16}}$           &  $\makecell[t]{0.36\\0.26}$  &   $\makecell[t]{3.52\\0.93}$    &  $\makecell[t]{4.53\\0.56}$  \\
\midrule
ESOL&         NO                                         & $\makecell[t]{0.002\\0.02\\0.003}$  &   $\makecell[t]{-5.48e^{-17}\\5.25e^{-16}}$           &  $\makecell[t]{0.25\\0.22}$  &   $\makecell[t]{4.11\\0.90}$    &  $\makecell[t]{4.87\\0.58}$  \\
\midrule
Lipophilicity&         YES                                         & $\makecell[t]{0.004\\0.02\\0.013}$  &   $\makecell[t]{-2.11e^{-17}\\ 6.15e^{-16}}$           &  $\makecell[t]{0.05\\0.05}$  &   $\makecell[t]{4.89\\0.24}$    &  $\makecell[t]{5.19\\0.22}$  \\

\midrule
BACE&         NO                                         & $\makecell[t]{0.007\\0.02\\0.022}$  &   $\makecell[t]{4.44e^{-19}\\9.03e^{-16}}$           &  $\makecell[t]{0.03\\0.02}$  &   $\makecell[t]{5.01\\0.19}$    &  $\makecell[t]{5.44\\0.23}$  \\

\midrule
BBBP&         NO                                         & $\makecell[t]{0.003\\0.03\\0.007}$  &  $\makecell[t]{-4.06e^{-17}\\7.61e^{-16}}$           &  $\makecell[t]{0.08\\0.10}$  &   $\makecell[t]{4.81\\0.56}$    &  $\makecell[t]{5.30\\0.38}$  \\

\midrule
ClinTox&         YES                                       & $\makecell[t]{0.003\\0.03\\0.008}$  &    $\makecell[t]{2.16e^{-17}\\6.72e^{-16}}$           &  $\makecell[t]{0.09\\0.14}$  &   $\makecell[t]{4.78\\0.62}$    &  $\makecell[t]{5.26\\0.38}$  \\
\end{tabular}
\end{table}

\subsection{Comparing T-Hop with the SOTA}
Looking back at Table \ref{tab:degen_vs_non_degen}, we see that the degenerate model performs better than the non-degenerate model on four out of the six datasets. It would therefore be instructive to compare the degenerate model with state-of-the-art models.
We display the comparison in Table \ref{tab:sota_2d}. The table shows that the degenerate model outperforms SOTA models on four out of the six datasets. Since the T-Hop framework does not incorporate 3-d geometry information, we limit the comparison in Table  \ref{tab:sota_2d} to only SOTA models that also do NOT incorporate 3-d geometry information. However, despite the fact that these SOTA models do not incorporate 3-d information, some of them nonetheless utilize sophisticated mechansims such as transformer-style aggregation as in GROVER \cite{grover} and AttentiveFP \cite{attentivefp}, whereas T-Hop does not. Moreover, some of the SOTA methods in Table \ref{tab:sota_2d}, such as DMPNN \cite{dmpnn}, also incorporate edge features whereas T-Hop does not. Based on the foregoing it becomes imperative to ask: why should such a simple degenerate model outperform more sophisticated models like GROVER \cite{grover}? Well one possible answer can be obtained from the GraphMixer paper \cite{graph_mixer} wherein their simple model was shown to outperform more sophisticated models, such as DySAT \cite{dynamic_dysat} and JODIE \cite{dynamic_jodie},  which incorporate transformer attention and RNNs. Hence, simpler models do not necessarily perform worse in all cases.

 \begin{table}
\footnotesize
\caption{Comparison of the degenerate model with state-of-the-art models limited to 2-d information }
\label{tab:sota_2d}
\centering
\begin{tabular}{ccccccc}
\toprule
 Metric& \multicolumn{3}{c} {ROC-AUC (Higher is better)} & \multicolumn{3}{c}{RMSE (Lower is better)}\\
\cmidrule(lr){2-4} \cmidrule(lr){5-7}
 Model& ClinTox & BBBP & BACE &FreeSolv& ESOL&Lipophilicity \\ 
\hline 
DMPNN&                   $90.6$  &  $ 71.0$   &  $80.9$         &  $2.082$  &  $1.050$  &  $\mathbf{0.683}$ \\ 
AttentiveFP&            $84.7$  &  $ 64.3$   &  $78.4$       &  $2.073$  &  $\mathbf{0.877}$  &  $0.721$ \\ 
N-GRAM$_{RF}$&    $77.5$  &  $ 69.7$    &  $77.9$        &  $2.688$  &  $1.074$  &  $0.812$ \\ 
N-GRAM$_{XGB}$&  $87.5                 $  &  $ 69.1$    &  $79.1$        &  $5.061$  &  $1.083$  &  $2.072$ \\ 
PretrainGNN&           $72.6$  &  $ 68.7$    &  $84.5$        &  $2.764$  &  $1.100$  &  $0.739$ \\ 
MolCLR&                   $       -                $  &  $ 72.2$    &  $82.4$        &  $2.594$  &  $1.271$  &  $          -                $ \\ 
Grover$_{base}$&  $81.2$  &  $ 70.0$     &  $82.6$       &  $2.176$  &  $0.983$  &  $0.817$ \\ 
Grover$_{large}$&  $76.2$  &  $ 69.5$     &  $81.0$       &  $2.272$  &  $0.895$  &  $0.823$ \\ 
T-Hop&                    $\mathbf{91.2}$  &  $ \mathbf{73.5}$      &  $\mathbf{86.4}$      &  $\mathbf{1.926}$  &  $0.898$  &  $0.737$ \\ 

\end{tabular}

\end{table}

\section{Conclusion}
We have presented a study on the usefulness of incorporating path information in molecular graphs for the task of predicting chemical properties in the arena of QSAR. We designed a framework termed T-Hop which allowed us to effectively achieve this task on six datasets from the MoleculeNet suite of datasets. T-Hop afforded us an opportunity to study the importance of path information due to how it can be toggled to operate in one of two modes: a non-degenerate mode which incorporates path information and a degenerate mode which does NOT incorporate any path information. Consequently, by comparing T-Hop's two modes on pertinent datasets we were able to evaluate the usefulness of path information on those datasets. Results showed that path information's usefulness varies from dataset to dataset, an observation that aligns with results from previous work.  But given that the usefulness of path information varies from dataset to dataset, we deemed it of value to explore avenues for predicting upfront whether or not path information would be useful on a given dataset. This is especially important because incorporating path information typically involves significantly extra computational overhead. Hence, we took the very first steps in this direction by building a classifier that can predict upfront whether or not path information would be useful on a given dataset. Finally, given that the degenerate version of our model performed better than the non-degenerate version in most cases, we went on to compare that degenerate version with SOTA models. To our surprise we found that, despite its simplicity, the degenerate model outperforms SOTA models in certain instances.  

\end{document}